\documentclass{mem}
\usepackage{natbib}\usepackage{txfonts}\usepackage{balance}
\usepackage{graphicx}
\usepackage[driverfallback=dvipdfm]{hyperref}
\idline{75}{282}
\begin{document}
\def\teff{$T\rm_{eff }$}
\def\kms{$\mathrm {km s}^{-1}$}

\title{AGB stars as tracers to IC\,1613 evolution}

   \subtitle{}

\author{
S.\ A.\ Hashemi\inst{1,2},
A.\,Javadi\inst{2},
J.\, Th.\,van Loon\inst{3},
}

\institute{
Physics Department, Sharif University of Technology, Tehran 1458889694, Iran
\and
School of Astronomy, Institute for Research in Fundamental Sciences (IPM),
    Tehran, 19395-5531, Iran
\and
Lennard-Jones Laboratories, Keele University, ST5 5BG, UK\\
\email{hashemi.seyedazim@gmail.com}
}

\authorrunning{S.A. Hashemi }

\titlerunning{IC\,1613 SFH}

\abstract{
we are going to apply AGB stars to find star formation history  for IC\,1613 galaxy;
this a new and simple method that works well for nearby galaxies. IC\,1613 is a Local
Group dwarf irregular galaxy that is located at distance of 750 kpc, a gas rich and 
isolated dwarf galaxy that has a low foreground extinction. We use the long period 
variable stars (LPVs) that represent the very final stage of evolution of stars 
with low and intermediate  mass at the AGB phase and are very luminous and cool so
that they emit maximum brightness in near--infrared bands. Thus near--infrared 
photometry with using stellar evolutionary models help us to convert brightness
to birth mass and age and from this drive star formation history of the galaxy. 
We will use the luminosity distribution of the LPVs to reconstruct the star formation
history--a method we have successfully applied in other Local Group galaxies. Our analysis
shows that the IC 1613 has had a  nearly constant star formation rate, without any dominant
star formation episode.
  
}
\maketitle{}

\section{Introduction}

IC\,1613 is an isolated irregular dwarf galaxy that is located in Local Group.
We adopt the mean distance of $∼ 750$ kpc ($( m - M)_0= 24 .37 \pm 0 .08$ mag ))
for this dwarf which is determined by Menzies et al.\ (2015) by fitting a 
period-luminosity relation to the C-rich Miras. Its proximity, low inclination
angle (i = 38$^{\circ}$) and low foreground reddening (E(B-V)= 0.025 mag; 
Menzies et al.\ 2015) make it very suitable target for studies of stellar 
population and evolution, interstellar medium and galaxy evolution.

Cool evolved stars are among the most accessible probes of stellar
populations due to their immense luminosity, from 2000 L$_\odot$ 
for tip-RGB stars, $\sim$ 10$^4$ L$_\odot$ for asymptotic giant branch
(AGB) stars, up to a few 10$^5$ L$_\odot$ for red supergiants
(Javadi et al.\ 2013). Their spectral energy distributions (SEDs) peak
around 1$\mu$m, so they stand out in the I-band (and reddening is 
reduced at long wavelengths). They have low surface gravity causing them
to pulsate radially on timescales of months to years. The most extreme 
examples among these long-period variables (LPVs) are Mira (AGB) variables,
which can reach amplitudes of ten magnitudes at visual wavelengths.
The variability helps identify these beacons; their luminosities
can be used to reconstruct the star formation history; and their
amplitudes pertain to the process by which they lose matter and
ultimately terminate their evolution.

Resolved stellar populations within galaxies allow us to derive star formation
histories on the basis of colour--magnitude diagram modelling, rather than from
integrated light. They also allow us to determine distances based on the tip of
the red giant branch (RGB), based on the period--luminosity relation of relatively
young populations of Cepheid variable stars, or based on the luminosities of old 
populations of RR Lyrae variable stars. Distances to unresolved galaxies are highly 
uncertain, especially in the local Universe where the Hubble flow does not yet 
dominate the peculiar motions of galaxies due to local density enhancements in
the cosmic web. However, stars can be resolved in all of the Local Group galaxies,
down to luminosities $<$ 1000 L$_\odot$ (Javadi et al.\ 2011a,b, 2015).

The Star Formation History (SFH) is one of the most important tracers of the
galaxies evolution. We have developed a novel method to use LPVs to reconstruct
the SFH (Javadi et al.\ 2011b,c, 2016, 2017; RezaeiKh et al.\ 2014; Golshan et al.\ 2017). 
In this paper we will use this new technique to represent the SFH of IC\,1613.

\section{Data}

We benefit from a number of published data sets in near--IR and mid--IR wavelengths
(see below). 

\subsection{The near-infrared data}
Menzies et al.\ (2015) published JHKs photometry from three years survey of the central
region of the IC\,1613 galaxy. They used Japanese--South African  
Infrared Survey Facility (IRSF) equipped  with SIRIUS camera.
They identified all objects  brighter than RGB--tip (K$\sim$18 mag) as supergiants or 
AGB stars (not foreground stars or background galaxies). Other data source in near--IR 
is from Sibbons et al.\ (2015). They used WFCAM camera on UKIRT to obtain JHK photometry 
of an area of 0.8 degree squares  centered on IC\,1613 galaxy. This survey is wider
and more complete than Menzies's work. From these data the iron abundance of 
$[Fe/H]=-1.26 \pm .08$ dex has been calculated and  their presented catalogue 
contains 843 AGB stars within 4.5 kpc of the IC\,1613 galactic center. 

\subsection{The mid--infrared data}

The Dust in Nearby Galaxies by {\it Spitzer} (DUSTiNGS) is a {\it Spitzer} Cycle 8 
program that imaged 50 nearby dwarf galaxies  in 3.6 and 4.5 $\mu$m bands  with 
a wide range in SFH and metallicity to detect dust--producing AGB stars (Boyer et al. 2015)
. The survey discovered,
50 new variable AGB candidates in IC\,1613, of which 34 are "extreme" (x--AGB) 
candidates. The red colors and variability of DUSTiNGS x--AGB candidates support 
the strong likelihood that these stars are true dust--producing AGB stars.

\begin{figure*}[t!]
\resizebox{\hsize}{!}{\includegraphics[clip=true]{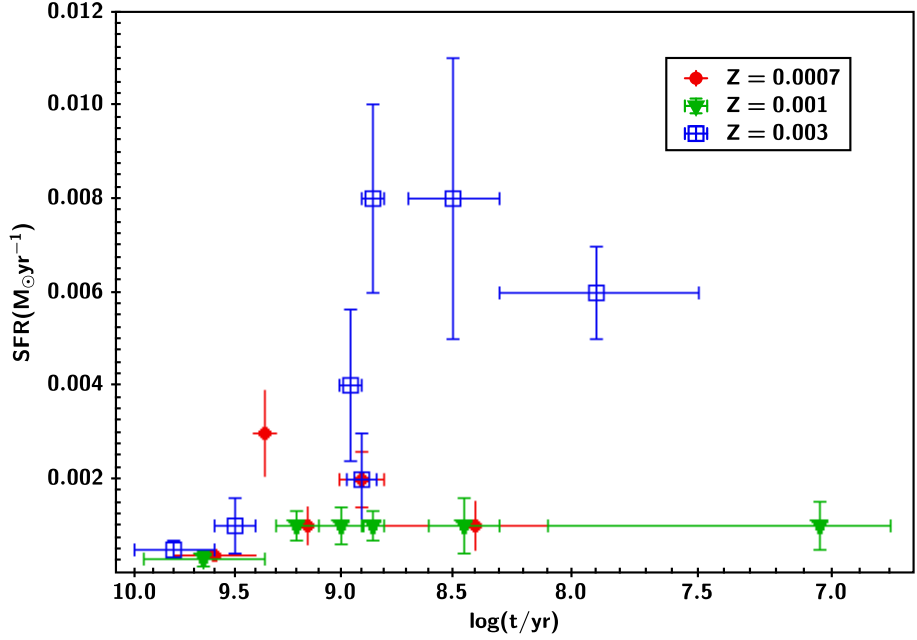}}
\caption{\footnotesize
 SFR versus look--back time (t) for three choices of  metallicity.
}
\label{fig1}
\end{figure*}

\begin{figure*}[t!]
\resizebox{\hsize}{!}{\includegraphics[clip=true]{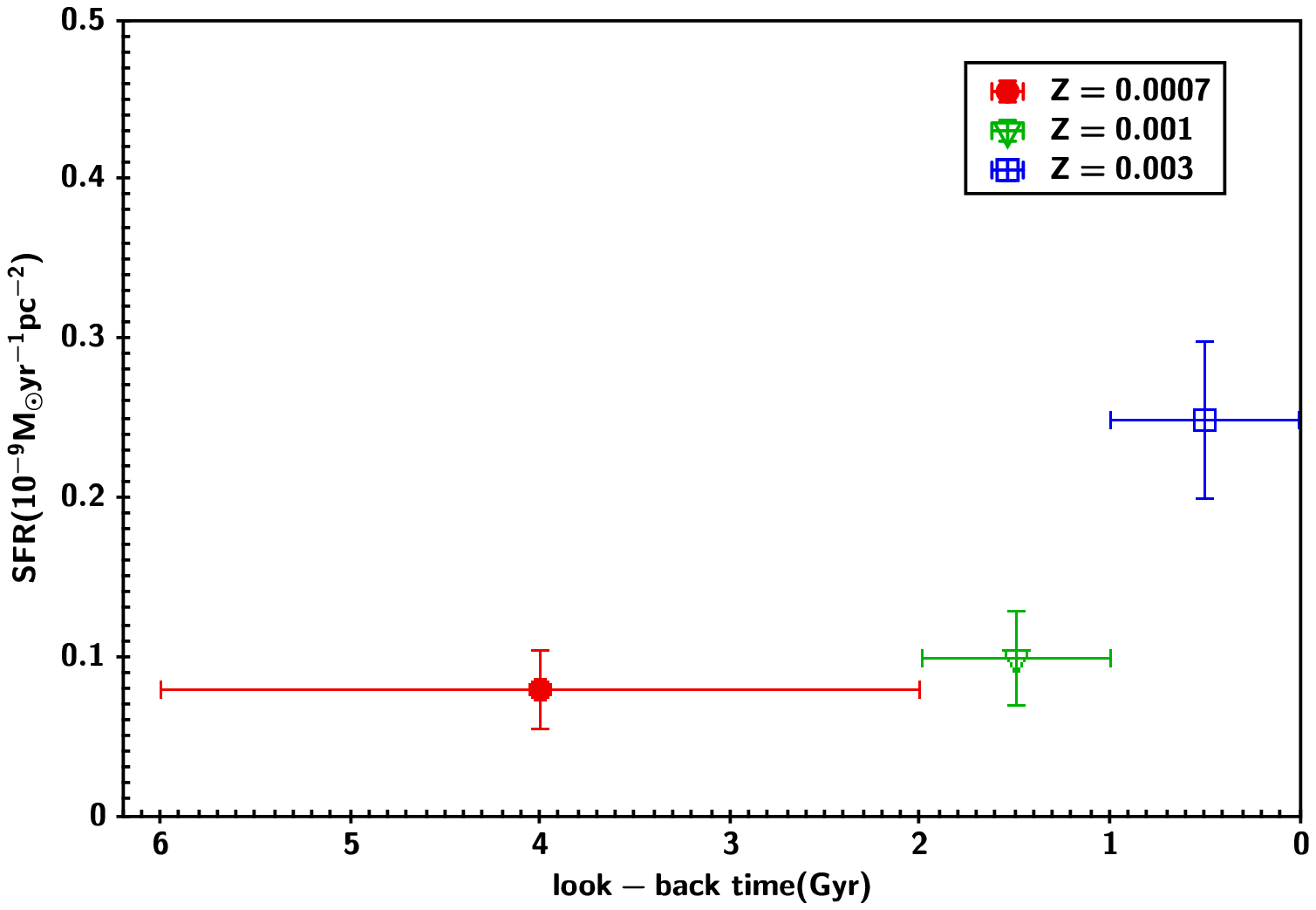}}
\caption{\footnotesize
 SFH with considering the metallicity evolution of the galaxy
}
\label{fig2}
\end{figure*}

\section{Star formation history}

The SFH of a galaxy is a measure of the rate at which the gas mass was converted
into stars over a time interval in the past. The
most evolved stars with low to intermediate mass, at the tip of the Asymptotic
Giant Branch (AGB) show brightness variations on timescales of $\approx100$ to
$>1000$ days due to radial pulsation. LPVs represent the most luminous phase
in their evolution, $\sim3000$--60,000 L$_\odot$, and reach their maximum
brightness at near-infrared wavelengths. Intermediate-mass AGB stars may
become carbon stars as a result of the dredge up of carbon synthesized in the
helium thermal pulses; the resulting change in opacity reddens their colours.
Since the maximum luminosity attained on the AGB relates to the star's birth
mass, we can use the brightness distribution function of LPVs to construct the
birth mass function and hence derive the Star Formation Rate (SFR) as a function of
time. In other words, the SFH is the 
SFR,
$\xi$ (in M$_\odot$ yr$^{-1}$), as a function of elapsed time, 
t. The amount of stellar
mass, $dM$, created during a time interval, $dt$, is:
\begin{equation}
dM(t) = \xi(t)\ dt.
\label{eq:eq1}
\end{equation}
Therefore, the number of formed stars are related to this mass by the following
equation:
\begin{equation}
dN(t) = \frac{\int_{\rm min}^{\rm max}f_{\rm IMF}(m)\ dm}
{\int_{\rm min}^{\rm max}f_{\rm IMF}(m)m\ dm}\ dM(t),
\label{eq:eq2}
\end{equation} 
where $f_{\rm IMF}$ is the Initial Mass Function (IMF). We use the IMF defined
in Kroupa (2001).
We need to relate this to the number of stars, $N$, which are variable at the
present time. If stars with mass between $m(t)$ and $m(t+dt)$ are LPVs at the
present time, then the number of LPVs created between times $t$ and $t+dt$ is:
\begin{equation}
dn(t) = \frac{\int_{m(t)}^{m(t+dt)}f_{\rm IMF}(m)\ dm}
{\int_{\rm min}^{\rm max}f_{\rm IMF}(m)\ dm}\ dN(t).
\label{eq:eq5}
\end{equation}
Substituting equation \ref{eq:eq1} and \ref{eq:eq2} in equation \ref{eq:eq5}
gives:
\begin{equation}
dn(t) = \frac{\int_{m(t)}^{m(t+dt)}f_{\rm IMF}(m)\ dm}
{\int_{\rm min}^{\rm max}f_{\rm IMF}(m)m\ dm}\ \xi(t)\ dt.
\label{eq:eq6}
\end{equation}

We are considering an age bin of $dt$, to determine $\xi(t)$. The number of
LPVs observed in this age bin, $dn^\prime$, depends on the duration of the
evolutionary stage during which the long period variability occurs:
\begin{equation}
dn^\prime(t) = \frac{\delta t}{dt}\ dn(t).
\label{eq:eq7}
\end{equation}

Finally, by combining the above equations we obtain a relation to calculate
the SFR based on LPVs  counts:
\begin{equation}
\xi(t) = \frac{\int_{\rm min}^{\rm max}f_{\rm IMF}(m)m\ dm}
{\int_{m(t)}^{m(t+dt)}f_{\rm IMF}(m)\ dm}\ \frac{dn^\prime(t)}{\delta t}.
\label{eq:eq8}
\end{equation}

To obtain the SFR we need to determine the individual stars' masses, ages
($t$) and duration of pulsation ($\delta t$). For this we rely on theoretical
models. The most appropriate theoretical models for our purpose are the Padova
models (Marigo et al.\ 2017)

\section{Results}
The SFR as a function of look-back time (t) in IC\,1613
 is shown in figure \ref{fig1}
 for different metallicites. The horizontal errorbars represent the age bins.

As a galaxy ages, the metallicity of the ISM – and hence
that of new generations of stars – changes as a result of
nucleosynthesis and feedback from dying stars. So we expect
older stars to have formed in more metal poor environments
than younger stars have. Figure \ref{fig2} shows the SFH when
we consider the effect of chemical
evolution of the galaxy.

\section{ conclusions}

\begin{itemize}
{\item Our analysis shows that the SFH of the observed field
in IC\,1613 is consistent with being  almost constant over the  lifetime 
of the galaxy.}

{\item Our results were obtained completely independently,
using different data and a different method, and yet they
are corroborated by previous work (Skillman et al.\ 2014).}
\end{itemize}

\bibliographystyle{aa}

\end{document}